\documentclass[pra,onecolumn,preprintnumbers,amsmath,amssymb]{revtex4}

\usepackage{bm,graphicx,multirow}

\newcommand{\vect}[1] {\ensuremath{\mathbf{#1}}} % display of N dim real vectors, or 3D vectors,
\newcommand{\transpose}{\ensuremath{^\text{\textsf{T}}}} % Symbol for transpose of vector or matrix

\newcommand{\pv}[2]{\ensuremath\begin{pmatrix} {#1} \\ {#2} \end{pmatrix}}  % displays column vector with the two given arguments as its components

\newcommand{\pmat}[4]{\ensuremath\begin{pmatrix} #1 & #2 \\ #3 & #4 \end{pmatrix} }  % general display of 2x2 matrix, with 4 arguments passed

\DeclareMathOperator{\sgn}{sgn}

\newcommand{\event}{\ensuremath{E}} % symbol for "event"
\newcommand{\mment}[2]{\ensuremath{\mathbf{#1}_{#2}}} % display of MEASUREMENTS (i.e. devices with particular settings, together with detectors arranged in some way).  Subscripted with #2.
\newcommand{\mmenta}[1]{\ensuremath{\mathbf{#1}}} % display of MEASUREMENTS (i.e. devices with particular settings, together with detectors arranged in some way).  Unsubscripted.

\newcommand{\cmment}[2] {\ensuremath{\widetilde{\mathbf{#1}}_{#2}}} % display of MEASUREMENT obtained by coarse-graining the detectors of another m'ment.  Subscripted with #2
\newcommand{\cmmenta}[1] {\ensuremath{\widetilde{\mmenta{#1}}}} % display of MEASUREMENT obtained by coarse-graining the detectors of another m'ment.  Unsubscripted.

\newcommand{\seq}[2]{\ensuremath{[#1,#2]}}   % symbolic representation of sequence of m'ments and their outcomes (2 measurements)

\newcommand{\lseq}[3]{\ensuremath{[#1,#2,#3]}}   % symbolic representation of sequence (3 measurements)

\newcommand{\lseqdots}[3]{\ensuremath{[#1,#2,#3,\dots]}}   % symbolic representation of sequence (3 measurements) with dots at end

\newcommand{\llseq}[4]{\ensuremath{[#1,#2,#3,#4]}}   % symbolic representation of sequence (4 measurements)

\newcommand{\weight}{pair}    % terminology: what we call the mathematical object attached to each sequence.

\newcommand{\weights}{pairs}  % plural of above

\newcommand{\wweights}{Pairs}  % capitalised form of \weights.

\newcommand{\sseq}[1]{\ensuremath{#1}}    % display of symbol for "sequence".
\newcommand{\pair}[1]{\ensuremath\vect{#1}}  % symbol for "number pair"
\newcommand{\cpair}[1]{\ensuremath({#1}_1, {#1}_2)^\textsf{T}} % a pair, written out component-wise.
\newcommand{\cvpair}[1]{\ensuremath\begin{pmatrix} #1_1 \\ #1_2 \end{pmatrix}}  % pair written out in column form
\newcommand{\lpair}[2]{\ensuremath{(#1, \,#2)\transpose}}  % pair with 2 arguments explicitly given written out with transpose superscript ("in-line pair")

\newcommand{\outcome}{m} % outcomes of measurements A_1, A_2, etc.

\newcommand{\bubble}[2]{\ensuremath(#1,#2)}   % symbol for an outcome which, when refined, only gives #1, #2 as possible outcomes (possible = non-zero probability).
\newcommand{\bbubble}[3]{\ensuremath(#1,#2,#3)}  % as above, but yields one of three possible outcomes when refined.

\newcommand{\pll}{\lor}  % symbol used for binary operation that combines two sequences "in parallel"

\newcommand{\ser}{\ensuremath{\mathop{\bm{\cdot}}}}  % symbol used for binary operation that combines two sequences "in series"

\newcommand{\wor}{\oplus}   % oplus, which eventually becomes (we discover) simply plus.
\newcommand{\wand}{\odot}    % otimes, which eventually becomes (we discover) simply times.
\newcommand{\op}{\circ}

\newcommand{\g}{\ensuremath\gamma} % shorthand for \gamma
\newcommand{\bg}{\ensuremath\bm{\gamma}}  % display of bold gamma

\newcommand{\vectf}{\ensuremath{\mathbf{F}}}  % display of vector function, F, symbol
\newcommand{\vectG}{\ensuremath{\mathbf{G}}}  % display of vector function, G, symbol

\newcommand{\vectg}[1]{\ensuremath{\mathbf{R}(#1)}}  % display of vector function, g, symbol w/arg
\newcommand{\vectgname}{\ensuremath{\mathbf{R}}}  % display of vector function, g, just the symbol

\newcommand{\prob}[1]{\ensuremath{P(\sseq{#1})}}   % display of p(A) = prob. associated with sequence A
\newcommand{\cprob}[1]{\ensuremath{p(\pair{#1})}}   % display of p(A) = prob. associated with sequence A

\newcommand{\hfunc}{\ensuremath{p}}   % display of p(a) = prob. associated with pair a

\newcommand{\revpair}[1]{\ensuremath\overleftarrow{\vect{#1}}}  % symbol for number pair that represents weight of a reverse sequence

\newcommand{\revsseq}[1]{\ensuremath\overleftarrow{#1}}    % display of symbol for a reverse sequence.

\newcommand{\tmatrix}{\pmat{\A}{\B}{\C}{\D}}  % "shear" transformation of pairs.

\newcommand{\A}{S}
\newcommand{\B}{T}
\newcommand{\C}{U}
\newcommand{\D}{V}

\begin{document}

\title{Origin of Complex Quantum Amplitudes and Feynman's Rules}

\author{Philip Goyal}
    \email{pgoyal@perimeterinstitute.ca}
    \affiliation{Perimeter Institute, Waterloo, Canada}
    \thanks{Corresponding author.}

\author{Kevin H.~Knuth}
    \email{kknuth@albany.edu}
    \affiliation{University at Albany~(SUNY), NY, USA}

\author{John Skilling}
    \email{skilling@eircom.net}
    \affiliation{Maximum Entropy Data Consultants Ltd, Kenmare, Ireland}

\begin{abstract}

Complex numbers are an intrinsic part of the mathematical formalism of quantum theory, and are perhaps its most characteristic feature.    In this paper, we show that the complex nature of the quantum formalism can be derived directly  from the assumption that  a \emph{pair} of real numbers is associated with each sequence of measurement outcomes, with the probability of this sequence being a real-valued function of this number pair.   By making use of elementary symmetry conditions, and without assuming that these real number pairs have any other algebraic structure, we show that these pairs must be manipulated according to the rules of \emph{complex} arithmetic.  We demonstrate that these complex numbers combine according to Feynman's sum and product rules, with the modulus-squared yielding the probability of a sequence of outcomes.

\end{abstract}

\maketitle

\section{Introduction}

Complex numbers are perhaps the most characteristic mathematical feature of quantum theory.  In recent years, there has been growing interest in elucidating the physical origin of this~(and other) mathematical features of quantum theory by deriving---or \emph{reconstructing}--- the quantum formalism from one or more physical principles,  and signiÞcant progress has been made~\cite{Wheeler89, Rovelli96, Zeilinger99, Fuchs02, Wootters80, Wootters-statistical-distance, Tikochinsky88a,  Brukner99, Summhammer94, Grinbaum03, Caticha98b, Clifton-Bub-Halvorson03, Goyal-QT2c, Goyal-QT, DAriano-operational-axioms, Hardy01a, MJWHall-Schroedinger-derivation, Reginatto-Schroedinger-derivation, Popescu-Rohrlich97}.   For example, many approaches are able to derive specific equations or predictions such as Schroedinger's equation or Malus' law~\cite{Wootters80, Wootters-statistical-distance, Brukner99, Summhammer94, MJWHall-Schroedinger-derivation, Reginatto-Schroedinger-derivation}.   However, the derivation of a significant part of the quantum formalism has thus far relied upon abstract assumptions such as the introduction of the complex number field~\cite{Grinbaum03, Clifton-Bub-Halvorson03} or upon several disparate features of quantum phenomena~\cite{DAriano-operational-axioms,Goyal-QT2c, Goyal-QT}.

In this paper, we present a novel reconstruction of Feynman's reformulation of quantum theory~\cite{Feynman48}.  Our approach differs from previous approaches in that it avoids \emph{ad hoc} introduction of the complex number field, and in that it rests essentially on a single postulate, namely:
\begin{quotation}
\noindent\textbf{\emph{Pair Postulate}}:~each sequence of measurement outcomes obtained in a given experiment is represented by a \emph{pair} of real numbers, where the probability associated with this sequence is a continuous, non-trivial function of both components of this real number pair.   
\end{quotation}
Since the probability is the only information which is accessible in a given experiment, this postulate expresses the simple idea that it requires twice as many degrees of freedom to describe a physical system than one can access through a given measurement.   This idea has played an important role in some previous attempts to reconstruct quantum theory~%
\footnote{For example, in \cite{Goyal-QT2c, Goyal-QT}, this idea is used to reconstruct the quantum formalism from a different point of view to that pursued here;   in~\cite{Stueckelberg60}, it is used in a partial reconstruction of quantum theory.  It has also been used in~\cite{Spekkens-toy-model} as the key idea to create a toy model of quantum theory.}%
, and can also be regarded as one way of stating Bohr's principle of complementarity~\cite{Bohr-complementarity}.

Using symmetry and consistency conditions that arise naturally in an operational framework, and making a few elementary physical assumptions, we show that this postulate leads to Feynman's rules of quantum theory.  Most importantly, we show that the number pairs assigned to each sequence of measurement outcomes must be manipulated according to the rules of \emph{complex} arithmetic, without assuming this at the outset.   Specifically, in the language in which Feynman's rules are usually expressed~\cite{Feynman48}, we show that, if the pair associated with a \emph{path} which a system classically can take from an initial event,~$\event_i$, to a final event,~$\event_f$, is written as a complex number, or \emph{amplitude,} then:
\begin{itemize} 
\item[(a)]If a system classically can take more than one path from~$\event_i$ to~$\event_f$, then the total amplitude for the transition is given by the \emph{sum} of the amplitudes associated with these paths,
\item[(b)]If the transition from~$\event_i$ to~$\event_f$ takes place via intermediate event~$\event_m$, the total amplitude  is given by the \emph{product} of the amplitudes for the transitions~$\event_i \rightarrow \event_m$ and~$\event_m \rightarrow \event_f$, and
\item[(c)]The \emph{probability} of the transition from~$\event_i$ to~$\event_f$ is proportional to the modulus-squared of the total amplitude for the transition.
\end{itemize}

Our approach is partly inspired by two previous reconstructions of Feynman's rules due to Tikochinsky~\cite{Tikochinsky88a} and Caticha~\cite{Caticha98b}.  Tikochinsky postulates that a complex number is associated with each path that a system can take between events.  By identifying a set of symmetries associated with these paths, he adapts an argument used by Cox to derive probability theory~\cite{Cox-PT-paper, Cox61} to show that these complex numbers must combine according to Feynman's rules.  Caticha's approach is similar, except that he operationalizes the classical notion of `path' within an experimental framework.  Both of these authors assume at the outset that complex arithmetic is to be used, and also implicitly assume that certain complex functions are analytic.  Such assumptions are given no \emph{a priori} physical justification, which detracts from the physical insight that these reconstructions can provide.    We show the complex structure of quantum theory need not be assumed, but can in fact be derived.  
The remainder of this paper is organized as follows.  

In Section~\ref{sec:experimental-framework}, we present an experimental framework which provides the basis for the reconstruction. The experimental framework provides a fully operational language which we use in place of the classical language of `paths' employed in Feynman's original formulation.    Sequences of measurement outcomes, and two operators that can be used to combine them in \emph{series} and in \emph{parallel,} are introduced.  Five fundamental symmetries associated with these operators are derived.  We then obtain a representation of this space of sequences by representing each sequence by a number pair and by requiring that these pairs combine through pair operators which share the same symmetries.

In Section~\ref{sec:derivation}, we use the symmetry constraints on the pair operators to determine their form.  This restricts their form to a few possibilities.  We then impose the requirement that the probability associated with a sequence is determined by the number pair that represents that sequence.  This requirement eliminates all but one of these possibilities, and completes the derivation by yielding the modulus-squared relation between the probability and the pair.   

We conclude in Section~\ref{sec:discussion} with a discussion of the results obtained and of potential future developments.

\section{Experimental Framework} \label{sec:experimental-framework}

We consider experimental set-ups in which a physical system is subject to successive measurements~$\mment{M}{1}, \mment{M}{2}, \mment{M}{3}, \dots$ at successive times~$t_1, t_2, t_3, \dots$.   
The system is allowed to undergo interactions in the intervening intervals.  We summarize the outcomes obtained in a given run of the experiment as a \emph{sequence}~$\sseq{A} = \lseqdots{\outcome_1}{\outcome_2}{\outcome_3}$.   
The measurements can be of different features of the system,
 but we shall label the outcomes of each measurement as~$1, 2, 3, 4, \dots$ as far as needed in each case.  

\begin{figure}
\includegraphics[width=6.75in]{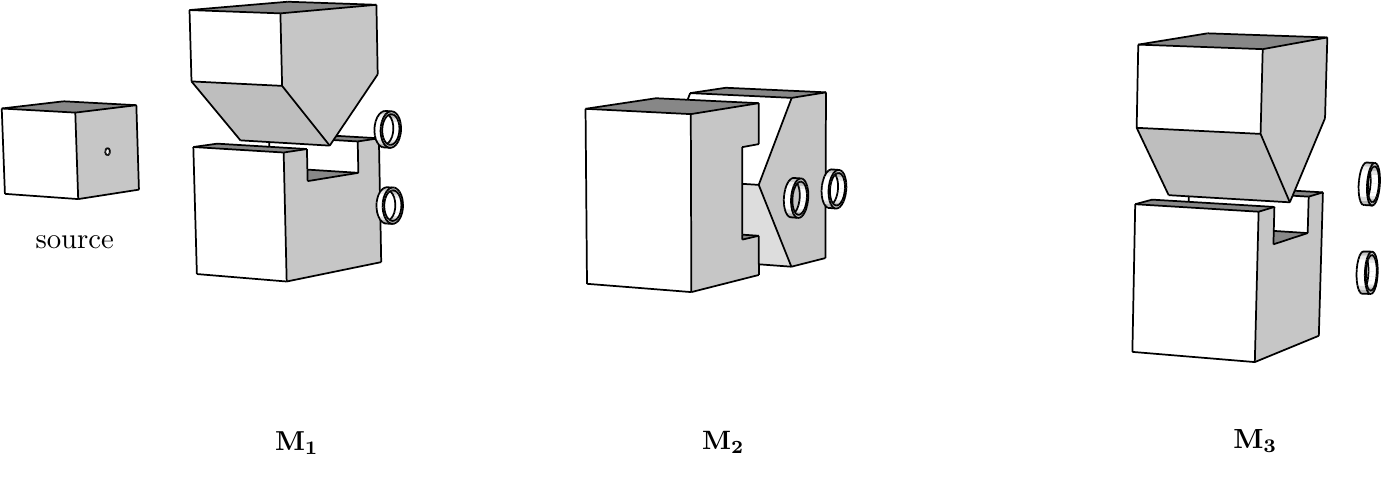}
\caption{\label{fig:SG-example} Schematic representation of a Stern-Gerlach experiment performed on silver atoms.  A silver atom from a source~(an evaporator) is subject to a sequence of measurements, each of which yields one of two possible outcomes registered by non-absorbing wire-loop detectors.  A run of the experiment yields outcomes~$\outcome_1, \outcome_2, \outcome_3$ of the measurements~$\mment{M}{1}, \mment{M}{2}, \mment{M}{3}$ performed at times~$t_1, t_2, t_3$, respectively.}
\end{figure}

Consider, for example, the Stern-Gerlach set-up shown in Fig.~\ref{fig:SG-example}. 
Here, a source supplies silver atoms which pass through the apparatus, undergoing successive measurements of components of spin.   Each measurement is performed by a magnet equipped with two wire-loop detectors (as sketched in the figure) which do not absorb the atoms.   Between the measurements, the spins may interact with a uniform magnetic field.  
For silver atoms, it is found experimentally that each measurement can only have two possible outcomes, which we label~$1$ and~$2$.    
These measurements are \emph{repeatable} in that the same result is always obtained if the same measurement is immediately repeated.

We might, for example, obtain the sequence~$\sseq{A} = \lseq{2}{1}{2}$, or perhaps~$\sseq{B} = \lseq{2}{2}{1}$.    Under repeated trials of this experiment, the probability distribution over the outcome of~$\mment{M}{3}$ is observed to be independent of any interactions the system had prior to~$\mment{M}{2}$, including the outcome of~$\mment{M}{1}$. 
In such a case, we say that the earlier measurement~$\mment{M}{2}$ \emph{establishes closure} with respect to the later~$\mment{M}{3}$~\footnote{See~Sec.~IIA of Ref.~\cite{Goyal-QT2c} for a fuller discussion of closure.}.  
Closure, in which current information overrides past information, is a basic feature of experiments on quantum systems.

\begin{figure} [!b]
\includegraphics[width=6.75in]{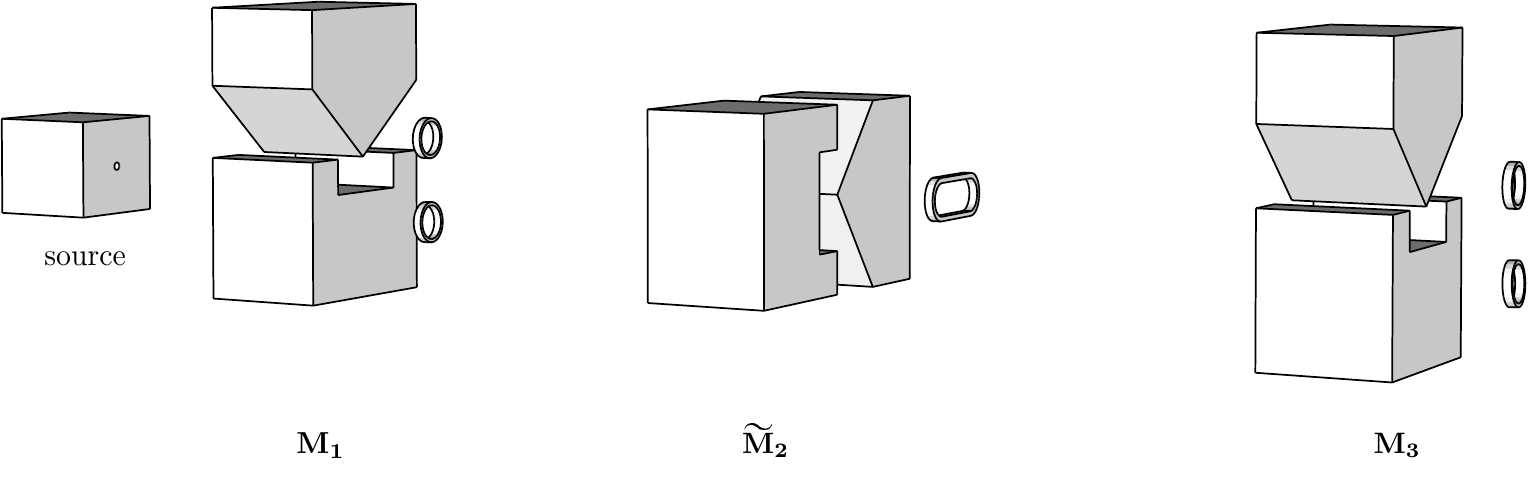}
\caption{\label{fig:SG-example2} A Stern-Gerlach experiment where the field of sensitivity of the intermediate measurement detector spans the fields of sensitivity of both of the detectors of the corresponding measurement in Fig.~\ref{fig:SG-example}.}
\end{figure}

We can also set up coarser experiments, such as the one shown in Fig.~\ref{fig:SG-example2}.  
Here, the measurement~$\cmment{M}{2}$ performed at~$t_2$ uses only a single detector whose field of sensitivity includes outcomes $1$ and $2$ of~$\mment{M}{2}$ in the original experiment.  
Now, if the coarser~$\cmment{M}{2}$ registers an atom, only outcome $1$ or $2$ could be obtained if
 measurement~$\mment{M}{2}$ was then performed immediately afterwards.   Accordingly, we write the outcome of $\cmment{M}{2}$ as~$(1,2)$, and we say that the measurement~$\cmment{M}{2}$ \emph{coarsens} outcomes~$1$ and~$2$ of the original~$\mment{M}{2}$.  
Using~$\mment{M}{2}$, outcome~$(1,2)$ can be refined to finer outcomes~$1$ and~$2$, but those latter outcomes cannot be further refined.  An outcome that cannot be further refined is said to be~\emph{atomic}.       If a measurement, such as~$\mment{M}{2}$, has all of its outcomes atomic, we shall call the measurement itself atomic.
The notation for non-atomic outcomes is naturally extended to the case where an outcome can be refined into more than two outcomes.

Generalizing the Stern-Gerlach example, we consider set-ups where the measurements (of a particular property) are repeatable and either atomic or coarsened versions of such,  and where the first and last measurements in the set-ups are atomic.
We also take the observed system to be sufficiently simple that the atomic measurements establishes closure with respect to any future measurement,  and that any interaction with the system between measurements preserves this closure.   

\subsection{Combining Sequences}

We now consider different ways in which sequences of measurement outcomes can be combined with one another to generate other sequences.    We use two kinds of relations between sequences, namely parallel and series combination.

\subsubsection{Sequences in Parallel}

\begin{figure}[!h]
\includegraphics[width=3.3in]{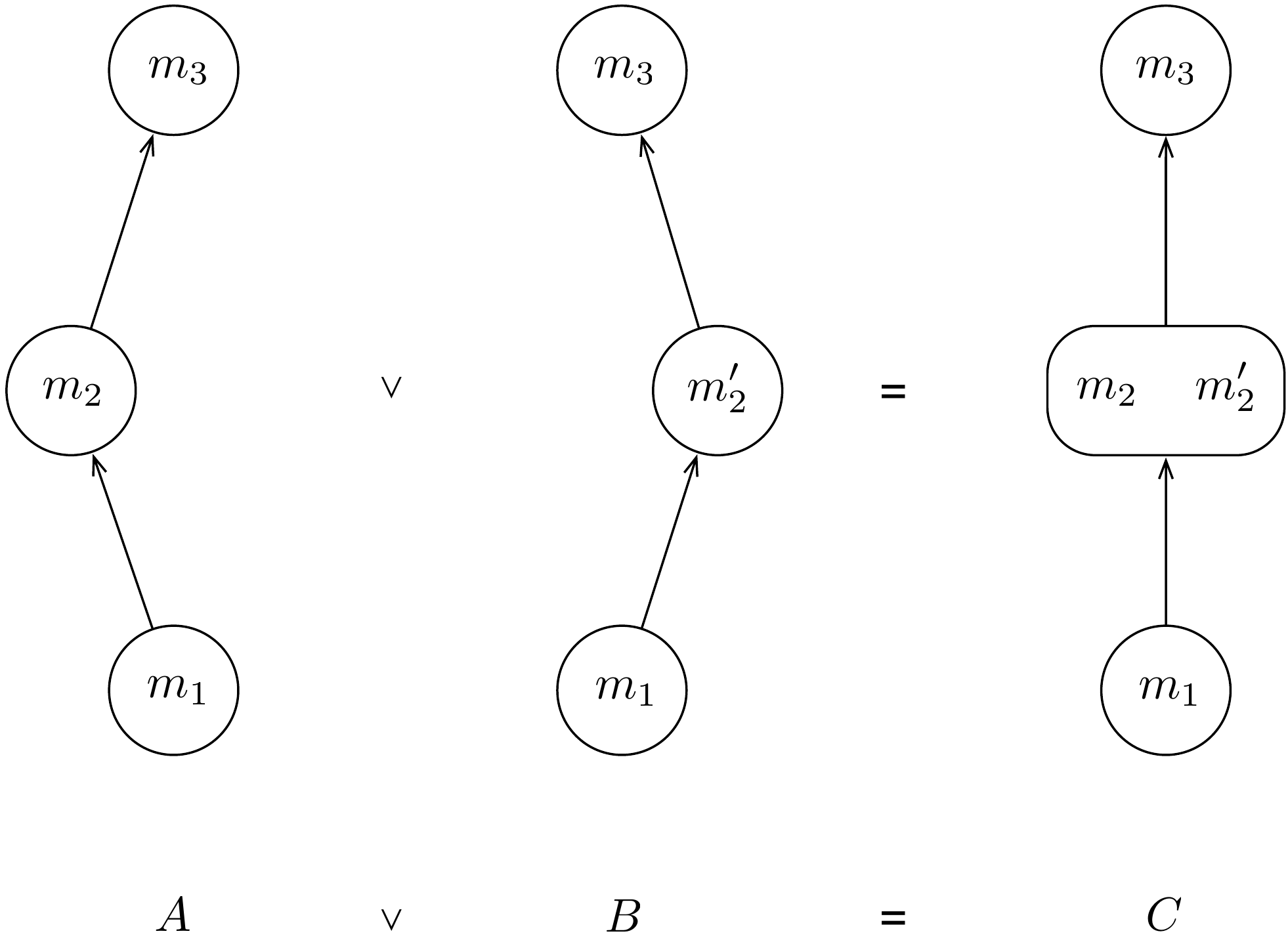}
\caption{\label{fig:seq-parallel} Combination of sequences in parallel.  Graphical depiction of the sequences~$\sseq{A} = \lseq{\outcome_1}{\outcome_2}{\outcome_3}, \sseq{B} = \lseq{\outcome_1}{\outcome_2'}{\outcome_3}$, and~$\sseq{C} = \lseq{\outcome_1}{\bubble{\outcome_2}{\outcome_2'}}{\outcome_3}$, respectively.}
\end{figure}

First, consider an experimental set-up consisting of three measurements,~$\mment{M}{1}, \mment{M}{2}$ and~$\mment{M}{3}$ performed in succession.  On one run, this generates sequence~$\sseq{A} = \lseq{\outcome_1}{\outcome_2}{\outcome_3}$ and, on another run, sequence~$\sseq{B} = \lseq{\outcome_1}{\outcome_2'}{\outcome_3}$, with~$\outcome_2 \neq \outcome_2'$.   Then consider a second set-up, identical to the first except that the intermediate measurement~$\cmment{M}{2}$ coarsens outcomes~$\outcome_2$ and~$\outcome_2'$ of~$\mment{M}{2}$, and suppose that this generates the sequence~$\sseq{C} = \lseq{\outcome_1}{\bubble{\outcome_2}{\outcome_2'}}{\outcome_3}$.
We shall say that the sequence~$\sseq{C}$ combines~$\sseq{A}$ and~$\sseq{B}$ \emph{in parallel}~(Fig.~\ref{fig:seq-parallel}).
We symbolize this relation by defining a binary operator,~$\pll$, which here acts on~$\sseq{A}$ and~$\sseq{B}$ to generate the sequence 
\begin{equation} \label{eqn:def-of-pll}
 \sseq{C} = \sseq{A} \pll \sseq{B}.
\end{equation}
Generally, the binary operator~$\pll$ combines any two sequences obtained  from the same experimental set-up differing in only one outcome.  

From the above definition, it follows at once that~$\pll$ is \emph{commutative} and~\emph{associative}.  To establish the first, notice that
\begin{equation*}
\sseq{B} \pll \sseq{A} 
		=   \lseq{\outcome_1}{\bubble{\outcome_2'}{\outcome_2}}{\outcome_{3}},
\end{equation*}
and since~${\bubble{\outcome_2}{\outcome_2'}} = {\bubble{\outcome_2'}{\outcome_2}}$, it follows that~$\pll$ is commutative,
\begin{equation} \label{eqn:parallel-commutativity} 
\sseq{A} \pll \sseq{B}   = \sseq{B} \pll \sseq{A}.
\end{equation}
To establish the second property, consider the three sequences~$\sseq{A} = \lseq{\outcome_1}{\outcome_2}{\outcome_3}, \sseq{B} = \lseq{\outcome_1}{\outcome_2'}{\outcome_3}$, and~$\sseq{C} = \lseq{\outcome_1}{\outcome_2''}{\outcome_3}$, with~$\outcome_2, \outcome_2'$ and~$\outcome_2''$ distinct.

These sequences can be combined  to form~$\sseq{D} = \lseq{\outcome_1}{\bbubble{\outcome_2}{\outcome_2'}{\outcome_2''}}{\outcome_3}$ in two different ways, namely
\begin{equation*}
\sseq{D} = (\sseq{A} \pll \sseq{B}) \pll \sseq{C}
\quad\quad\text{and}\quad\quad
\sseq{D} = \sseq{A} \pll (\sseq{B} \pll \sseq{C}),
\end{equation*}
which implies that~$\pll$ is associative,
\begin{equation}   \label{eqn:parallel-associativity} 
(\sseq{A} \pll \sseq{B}) \pll \sseq{C}    = \sseq{A}  \pll (\sseq{B} \pll \sseq{C}).
\end{equation}

\subsubsection{Sequences in Series}

\begin{figure}[!h]
\includegraphics[width=2.75in]{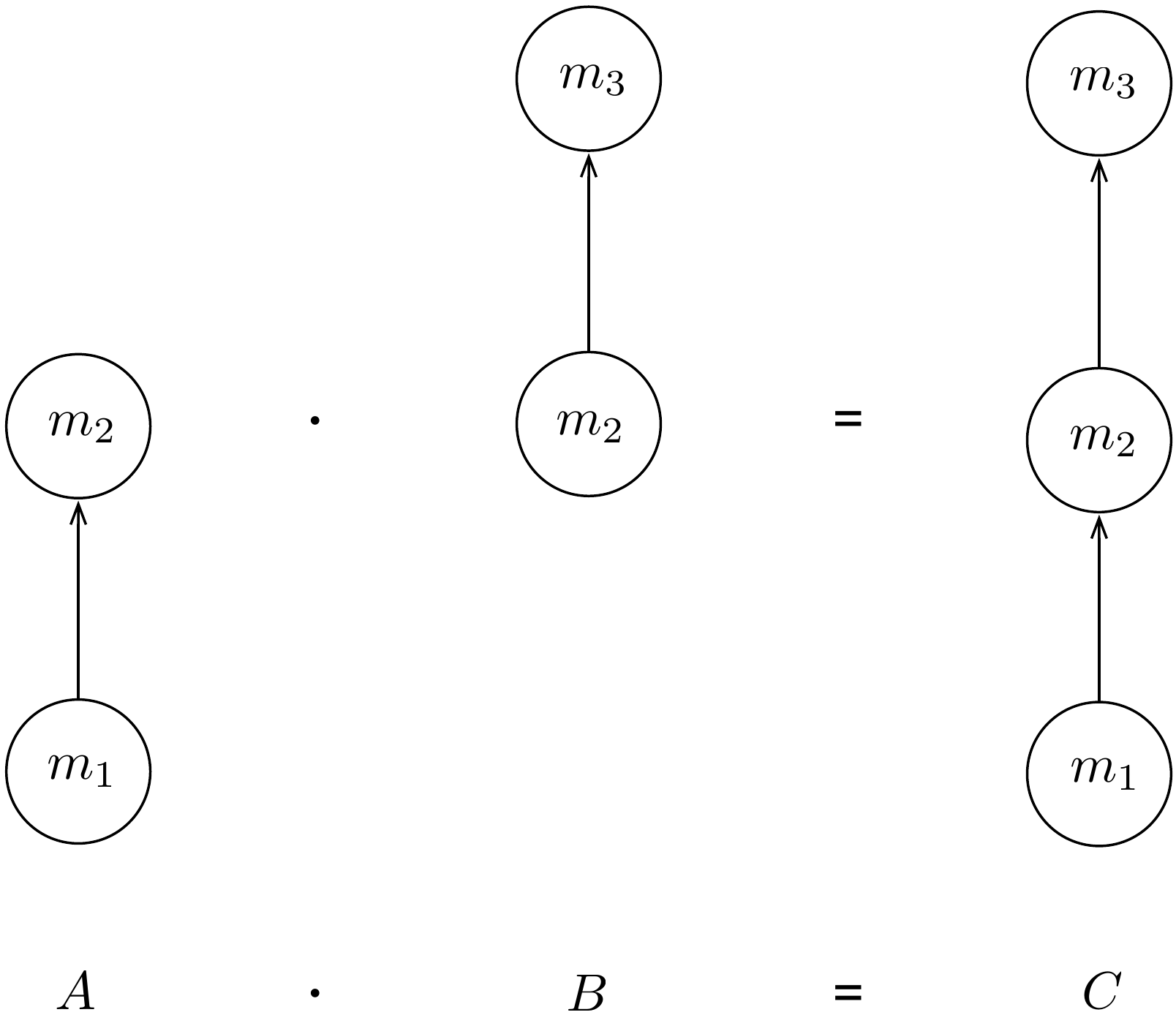}
\caption{\label{fig:seq-series}Combination of sequences in series. Graphical depiction of sequences~$\sseq{A} = \seq{\outcome_1}{\outcome_2}, \sseq{B} = \seq{\outcome_2}{\outcome_3}$ and~$\sseq{C} = \lseq{\outcome_1}{\outcome_2}{\outcome_3}$, respectively.}
\end{figure}

Consider the two sequences~$\sseq{A} = \seq{\outcome_1}{\outcome_2}$ and~$\sseq{B} = \seq{\outcome_2}{\outcome_3}$, in which outcome~$\outcome_2$ is the same in each~(see Fig.~\ref{fig:seq-series}).    
We now define the binary operator,~$\ser$\,, which chains two such sequences in series.  
This acts on~$\sseq{A}$ and~$\sseq{B}$ to generate the sequence
\begin{equation} \label{eqn:def-of-ser}
\sseq{C} =  \sseq{A} \ser \sseq{B} 
			=  \lseq{\outcome_1}{\outcome_2}{\outcome_3}.
\end{equation}
Generally, the binary operator~$\,\ser\,$ combines together any two sequences obtained from experimental set-ups where the last measurement~(and the outcome) of one sequence coincides with the first measurement~(and the outcome) of the other.

By considering the three sequences~$\sseq{A} = \seq{\outcome_1}{\outcome_2}$, $\sseq{B} = \seq{\outcome_2}{\outcome_3}$, and~$\sseq{C} = \seq{\outcome_3}{\outcome_4}$, we see that~$\ser$ is associative,
\begin{equation} \label{eqn:series-associativity} 
(\sseq{A} \ser \sseq{B}) \ser \sseq{C}    = \sseq{A}  \ser \, (\sseq{B} \ser \sseq{C}).
\end{equation}
Finally, consider the sequences~$\sseq{A} = \lseq{\outcome_1}{\outcome_2}{\outcome_3}$ and~$\sseq{B} = \lseq{\outcome_1}{\outcome_2'}{\outcome_3}$ and~$\sseq{C} = \seq{\outcome_3}{\outcome_4}$.   These can be combined in two equivalent ways to yield~$\sseq{D} = \llseq{\outcome_1}{\bubble{\outcome_2}{\outcome_2'}}{\outcome_3}{\outcome_4}$, namely
\begin{equation*}
\sseq{D} = (\sseq{A} \pll \sseq{B}) \ser \sseq{C}
\quad\quad\text{and}\quad\quad
\sseq{D} = (\sseq{A} \ser \sseq{C}) \pll (\sseq{B} \ser \sseq{C}).
\end{equation*}
Hence, the operation~$\ser$ is right-distributive over~$\pll$,
\begin{equation} \label{eqn:right-distributivity} 
 (\sseq{A} \pll \sseq{B}) \ser \sseq{C} = (\sseq{A} \ser \sseq{C}) \pll (\sseq{B} \ser \sseq{C}).
\end{equation}
Similar considerations show that~$\ser$ is also left-distributive over~$\pll$,
\begin{equation} \label{eqn:left-distributivity}
\sseq{C} \ser\, (\sseq{A} \pll \sseq{B}) = (\sseq{C} \ser \sseq{A}) \pll (\sseq{C} \ser \sseq{B}).
\end{equation}

\subsection{Sequence \wweights} \label{sec:weights-of-sequences}

Following the first part of our pair postulate, we represent each sequence,~$\sseq{A}$, by a real number~\weight,~$\pair{a} = \cpair{a}$.   We have determined that the parallel and series operators,~$\pll$ and~$\ser$, possess the symmetries given in Eqs.~\eqref{eqn:parallel-commutativity}, \eqref{eqn:parallel-associativity}, \eqref{eqn:series-associativity}, \eqref{eqn:right-distributivity}, and~\eqref{eqn:left-distributivity}.  These symmetries must be reflected in the representation.   For example, if pairs~$\pair{a}, \pair{b}$ represent the sequences~$\sseq{A}, \sseq{B}$, respectively, then the pair~$\pair{c}$ that represents~$\sseq{C} = \sseq{A}\pll\sseq{B}$ must be determined by~$\pair{a}, \pair{b}$ through the relation
\begin{equation} \label{eqn:def-of-wor}
\pair{c} = \pair{a} \wor \pair{b},
\end{equation}
where~$\wor$ is a pair-valued binary operator, assumed continuous, to be determined.  Then, since~$\pll$ is commutative, we have~$\sseq{A}\pll\sseq{B} = \sseq{B}\pll\sseq{A}$, so that
\begin{equation}
\pair{a} \wor \pair{b} 				= \pair{b} \wor \pair{a}.   							\tag{S1} 	\label{eqn:wor-commutativity} 
\end{equation}
In addition, since~$\pll$ is associative,
\begin{equation}
(\pair{a} \wor \pair{b} ) \wor \pair{c}	 = \pair{a} \wor (\pair{b}  \wor \pair{c}).	 \tag{S2}  \label{eqn:wor-associativity}
\end{equation}

Similarly, if the sequences~$\sseq{A}, \sseq{B}$ and~$\sseq{C}$ are related by~$\sseq{C} = \sseq{A} \ser \sseq{B}$, then the pair~$\pair{c}$ that represents~$\sseq{C}$ must be determined by~$\pair{a}, \pair{b}$ through the relation
\begin{equation} \label{eqn:def-of-wand}
\pair{c} = \pair{a} \wand \pair{b},
\end{equation}
where~$\wand$ is another pair-valued binary operator, assumed continuous, also to be determined. From the associativity of~$\ser\,$, it follows that~$\wand$ also has associative symmetry,
\begin{equation}
(\pair{a} \wand \pair{b} ) \wand \pair{c}	= \pair{a} \wand (\pair{b}  \wand \pair{c}).   \tag{S3}  \label{eqn:wand-associativity}
\end{equation}
Finally, since~$\,\ser\,$ is right- and left-distributive over~$\pll$, it follows that the pair operators also have distributive symmetry,
\begin{align}
(\pair{a} \wor \pair{b} ) \wand \pair{c}	&= (\pair{a} \wand \pair{c})  \wor (\pair{b} \wand \pair{c})  \tag{S4} \label{eqn:wand-distributivity1}  \\
\pair{a} \wand (\pair{b}  \wor \pair{c}) 	&= (\pair{a} \wand \pair{b})  \wor (\pair{a} \wand \pair{c}). \tag{S5} \label{eqn:wand-distributivity2}
\end{align}

\section{Derivation of Feynman's Rules}
\label{sec:derivation}

In Sec.~\ref{sec:derivation1}, we shall use the symmetry equations~\eqref{eqn:wor-commutativity}--\eqref{eqn:wand-distributivity2} to fix the form of~$\oplus$ and to restrict~$\odot$ to one of five possible forms.  Then, in Sec.~\ref{sec:derivation2}, we shall introduce a connection between pairs and probabilities.  This will restrict~$\odot$ to a unique form, and fix the functional connection between pairs and probabilities.

\subsection{Solution of the Symmetry Equations for~$\wor$ and~$\wand$}
\label{sec:derivation1}

\subsubsection{Solution of Commutativity and Associativity Equations for~$\wor$}

Commutativity and associativity of~$\wor$ impose 
strong constraints on the possible forms that the operator can take.  To illustrate the nature of this constraint, consider a binary operator~$\op$ which acts over the real numbers.  In this one-dimensional case, there exist a number of theorems which show that, if operator~$\op$ is continuous and associative, and possesses a small number of additional properties~\footnote{%
For example, Acz\'{e}l~\cite{Aczel-associativity} shows that the additional property of cancellativity suffices, namely that, in general,~$x_1 \op y = x_2 \op y$ implies~$x_1 = x_2$ and, similarly,~$x \op y_1 = x \op y_2$ implies~$y_1 = y_2$.
},
then the operator must satisfy the equation
\begin{equation} \label{eqn:soln-of-real-associativity}
f(x \op y) =  f(x) + f(y),
\end{equation}
where~$f$ is a continuous and strictly monotonic function.
That is, given a binary operator over the reals satisfying the above-mentioned conditions, one can always invertibly transform the real-line such that, in the transformed space, the operator~$\op$ is represented by the addition operator.  Hence, without any loss of generality, one can choose to perform the composition operation in the transformed space.  Parenthetically, this result forms the basis of Cox's derivation of probability theory~\cite{Cox-PT-paper, Cox61} and is the rationale for additivity in measure theory~\cite{Knuth-laws-ordering-relations2003}.

In the two-dimensional case with which we are concerned here, an analogous result holds, namely that, for continuous, associative and commutative~$\wor$, then the operator must satisfy the equation
\begin{equation}  \label{eqn:soln-of-wor-associativity}
\vectf(\pair{a} \wor \pair{b}) =  \vectf(\pair{a}) + \vectf(\pair{b}),
\end{equation}
where~$\vectf$ is an invertible and continuous pair-valued function.
This was proved by Acz\'{e}l and Hossz\'{u}~\cite{Aczel-Hosszu} with the aid of minor technical assumptions.

Hence, without any loss of generality, we can transform the space of \weights\ such that the operator~$\wor$ becomes represented in standard form by the additive operator.  Explicitly, in this standard form,
\begin{equation}  \label{eqn:weight-sum-rule}
\cvpair{a} \wor \cvpair{b} = \pv{a_1 + b_1}{a_2 + b_2},
\end{equation}
which we refer to as the \emph{sum rule}.    Note that the only freedom left in the sum rule is a real invertible linear transformation of the space of pairs,
\begin{equation}  \label{eqn:def-of-regrading}
\begin{aligned}
	\begin{pmatrix}
	x_1' \\ x_2'
	\end{pmatrix}
		&=  \tmatrix
		\begin{pmatrix}
		x_1 \\ x_2
		\end{pmatrix},
\end{aligned}
\end{equation}
with~$\A\D-\B\C \neq 0$.   That is, whenever the sum rule holds between pairs, it also holds between the corresponding transformed pairs.  We shall make use of this fact below.

\subsubsection{Solution of Associativity and Distributivity Equations for~$\wand$}  \label{sec:soln-wand-eqns}

Having shown that~$\wor$ corresponds to component-wise addition of number pairs, we proceed to show that~$\wand$ corresponds to a form of multiplication.

\paragraph{Distributivity of~$\wand$.}
First, define
\begin{equation*}
\vectG(\pair{a}, \pair{b}) = \pair{a}\wand\pair{b},
\end{equation*}
where the pair-valued function~$\vectG$ is to be determined through Eqs.~\eqref{eqn:wand-distributivity1} and~\eqref{eqn:wand-distributivity2}, which become
\begin{equation*}
\begin{aligned}
\vectG(\pair{a} + \pair{b}, \pair{c}) &= \vectG(\pair{a}, \pair{c}) + \vectG(\pair{b}, \pair{c}) \\
\vectG(\pair{a}, \pair{b} + \pair{c}) &= \vectG(\pair{a}, \pair{b}) + \vectG(\pair{a}, \pair{c}).
\end{aligned}
\end{equation*}
Defining $r\pair{a} = r\cpair{a} \equiv \cpair{ra}$,  in accordance with Eq.~\eqref{eqn:weight-sum-rule}, with $r$ real, it follows that
\begin{equation*}
\vectG(r_1 \pair{a}, r_2 \pair{b}) = r_1 r_2 \vectG(\pair{a}, \pair{b}).
\end{equation*}
Introducing two-dimensional basis pairs~$\pair{e}_1$ and~$\pair{e}_2$, it then follows that
\begin{equation*}
\begin{aligned}
\vectG(\pair{a}, \pair{b}) 	&= \vectG(a_1 \pair{e}_1 + a_2  \pair{e}_2, b_1 \pair{e}_1 + b_2  \pair{e}_2)  \\
					&= a_1b_1\vectG(\pair{e}_1, \pair{e}_1) + a_1b_2 \vectG(\pair{e}_1, \pair{e}_2)
						+ a_2b_1 \vectG(\pair{e}_2, \pair{e}_1) + a_2b_2 \vectG(\pair{e}_2, \pair{e}_2)  \\
					&= a_1b_1 \pv{\g_1}{\g_5}  + a_1b_2 \pv{\g_2}{ \g_6} + a_2b_1 \pv{\g_3}{\g_7}  
						+ a_2b_2 \pv{\g_4}{\g_8},
\end{aligned}
\end{equation*}
where~$\bg = (\g_1, \g_2, \g_3, \g_4; \g_5, \g_6, \g_7, \g_8)$ is a real-valued vector to be determined, in which the semicolon partitions~$\bg$ into components that, respectively, effect the first and second part of the real pair.     Hence, the left- and right-distributivity of~$\wand$ over~$\wor$ implies that~$\pair{a}\wand\pair{b}$ has the bilinear multiplicative form
\begin{equation} \label{eqn:wand-bilinear-form}
\cvpair{a} \wand \cvpair{b} = \pv{\g_1 a_1b_1 + \g_2a_1b_2 + \g_3 a_2b_1 + \g_4 a_2 b_2}{\g_5 a_1b_1 + \g_6 a_1b_2 + \g_7 a_2b_1 + \g_8 a_2 b_2}.
\end{equation}

\paragraph{Associativity of~$\wand$.}
Substituting this form of~$\pair{a}\wand\pair{b}$ into the $\wand$-associativity condition, Eq.~\eqref{eqn:wand-associativity}, and solving the resulting equations~(see Appendix~\ref{app:wand-associativity}), one finds that~$\bg$ can take one of three possible forms, namely a commutative form
\begin{equation}  \label{eqn:gamma-commutative}
\bg = (\theta - \psi\epsilon,\, \phi\epsilon,\, \phi\epsilon,\, \phi;\  \theta\epsilon,\, \theta,\, \theta,\, \psi + \phi\epsilon),
\end{equation}
with real constants~$\theta, \phi, \psi, \epsilon$, and two non-commutative forms
\begin{align} 
\bg 	&=	 (\theta,\, \phi,\, 0,\, 0;\,0,\, 0,\, \theta,\, \phi)  \label{eqn:gamma-noncommutative1}   \\ 
\bg 	&=	 (\theta,\, 0,\, \psi,\, 0;\, 0,\, \theta,\, 0,\, \psi).  \label{eqn:gamma-noncommutative2}   
\end{align}

Using the freedom described in Eq.~\eqref{eqn:def-of-regrading}, we can transform these solutions to standard forms.  To do so, we note that, under the transformation of~Eq.~\eqref{eqn:def-of-regrading}, the relation~$\pair{c} = \pair{a} \wand \pair{b}$ transforms to
\begin{equation*}
\begin{aligned}
\cvpair{c'}	&= 	 \tmatrix \pv{\g_1 a_1b_1 + \g_2a_1b_2 + \g_3 a_2b_1 + \g_4 a_2 b_2}{\g_5 a_1b_1 + \g_6 a_1b_2 + \g_7 a_2b_1 + \g_8 a_2 b_2}   \\
		&= \pv{\g_1' a_1'b_1' + \g_2'a_1'b_2' + \g_3' a_2'b_1' + \g_4' a_2' b_2'}{\g_5' a_1'b_1' + \g_6' a_1'b_2' + \g_7' a_2'b_1' + \g_8' a_2' b_2'},
\end{aligned}
\end{equation*}
where
\begin{equation}
\cvpair{a'} = \tmatrix \cvpair{a}
\quad\quad\text{and}\quad\quad
\cvpair{b'} = \tmatrix \cvpair{b}
\quad\quad\text{and}\quad\quad
\cvpair{c'} = \tmatrix \cvpair{c},
\end{equation}
and where~$\bg' = (\g_1', \dots, \g_8')$ is the representation of~$\bg$ in the space of the transformed pairs.  
Equating coefficients of~$a_1', a_2', b_1', b_2'$ identifies~$\bg'$ as
\begin{equation}  \label{eqn:gamma-transformation}
	\begin{pmatrix}
	\g_1'\\ \g_2' \\ \g_3' \\ \g_4' \\ \g_5' \\ \g_6' \\ \g_7' \\ \g_8' 
	\end{pmatrix} 
	 = \frac{1}{\A\D - \B\C}
	 \begin{pmatrix}
		 \A^2\D &   \A\C\D &   \A\C\D &  \C^2\D & -\A^2\B &  -\A\B\C &  -\A\B\C & -\B\C^2 \\
                  \A\B\D &  \A\D^2 &   \B\C\D &  \C\D^2 & -\A\B^2 &  -\A\B\D & -\B^2\C &  -\B\C\D \\
                  \A\B\D &   \B\C\D &  \A\D^2 &  \C\D^2 & -\A\B^2 & -\B^2\C &  -\A\B\D &  -\B\C\D \\
                 \B^2\D &  \B\D^2 &  \B\D^2 &   \D^3 &  -\B^3 & -\B^2\D & -\B^2\D & -\B\D^2 \\
                -\A^2\C & -\A\C^2 & -\A\C^2 &  -\C^3 &   \A^3 &  \A^2\C &  \A^2\C &  \A\C^2 \\
                 -\A\B\C &  -\A\C\D & -\B\C^2 & -\C^2\D &  \A^2\B &  \A^2\D &   \A\B\C &   \A\C\D \\
                 -\A\B\C & -\B\C^2 &  -\A\C\D & -\C^2\D &  \A^2\B &   \A\B\C &  \A^2\D &   \A\C\D \\
                -\B^2\C &  -\B\C\D &  -\B\C\D & -\C\D^2 &  \A\B^2 &   \A\B\D &   \A\B\D &  \A\D^2 
	\end{pmatrix}
	\begin{pmatrix}
	\g_1 \\ \g_2 \\ \g_3 \\ \g_4 \\ \g_5 \\ \g_6 \\ \g_7 \\ \g_8
	\end{pmatrix}.
\end{equation}
Using this transformation, Eqs.~\eqref{eqn:gamma-commutative}, \eqref{eqn:gamma-noncommutative1} and~\eqref{eqn:gamma-noncommutative2} can be reduced to standard forms.

In particular, Eq.~\eqref{eqn:gamma-commutative} takes the standard form
\begin{equation}  \label{eqn:gamma-commutative-standard}
\bg = (1,\,0,\,0,\,\mu; \,0,\,1,\,1,\,0),
\end{equation}
where~$\mu = \sgn(4\theta\phi + \psi^2)$ can be~$-1$, $0$, or~$+1$.   
Through Eq.~\eqref{eqn:wand-bilinear-form}, case~$\mu=-1$ gives what we recognize as complex multiplication, while the cases~$\mu =0$ and~$\mu =+1$ give variations thereof.  
The transformation needed to recover Eq.~\eqref{eqn:gamma-commutative} from this standard form is
\begin{equation*}
	\tmatrix
	=
	\frac{1}{2}
	\begin{pmatrix}
	2\theta - \psi\epsilon 	& 2\phi\epsilon + \psi \\
	\epsilon\Delta 			& \Delta
	\end{pmatrix},
\end{equation*}
where
\begin{equation*}
\Delta = \begin{cases}
		\sqrt{|4\theta\phi + \psi^2|} 	&	\text{if}~\mu = \pm 1  \\
		1								&	\text{if}~\mu = 0.
		\end{cases}
\end{equation*}
When $SV-TU \ne 0$, the inverse of this transformation exists, so that Eq.~\eqref{eqn:gamma-commutative} can be returned to the standard form, Eq.~\eqref{eqn:gamma-commutative-standard}.
Note that this standard form with~$\mu=+1$ can be reached from the even simpler form
\begin{equation} \label{eqn:complex3-separable}
\bg = (1,\,0,\,0,\,0;\,0,\,0,\,0,\,1),
\end{equation}
that we use later, by applying the invertible transformation
\begin{equation*}
    \tmatrix
	= 
	\begin{pmatrix}
	1 		&	-1 \\
	1 		&	\,\,\,\,1
    \end{pmatrix}.
\end{equation*}

When, on the other hand, $SV-TU = 0$, the transformation would be singular, hence disallowed.
This would happen if $\theta = \zeta\epsilon$ where $\zeta = \psi + \phi\epsilon$.
Eq.~\eqref{eqn:gamma-commutative} would then have been
\[
\bg = (\phi\epsilon^2,\, \phi\epsilon,\, \phi\epsilon,\, \phi;\  \zeta\epsilon^2,\, \zeta\epsilon,\, \zeta\epsilon,\, \zeta),
\]
Observing the linear relation $\zeta c_1 = \phi c_2$ between the components of any product $\pair{c} = \pair{a}\wand\pair{b}$ thus defined,
 we note that this could be transformed by rotation to $\pair{c}' = \lpair{c_1'}{0}$.
This lacks the two components that we demand of an arbitrary pair, so the singular case is inadmissible.

Continuing in this style, Eq.~\eqref{eqn:gamma-noncommutative1} takes the standard form
\begin{equation}  \label{eqn:gamma-noncommutative1-standard}
\bg = (1,0,0,0;0,0,1,0).
\end{equation}
The transformation needed to recover Eq.~\eqref{eqn:gamma-noncommutative1} from this standard form is
\begin{equation*}
	\tmatrix	= 
	\begin{pmatrix}
	\theta 		&		\phi \\   
	-\phi 		&		\theta    
    \end{pmatrix},
\end{equation*}
which is invertible unless $\theta$ and $\phi$ both vanish.  
Similarly, the other non-commutative form  Eq.~\eqref{eqn:gamma-noncommutative2} has the standard form
\begin{equation}  \label{eqn:gamma-noncommutative2-standard}
\bg = (1,0,0,0; 0,1,0,0).
\end{equation}
This transforms to Eq.~\eqref{eqn:gamma-noncommutative2} through
\begin{equation*}
	\tmatrix	= 
	\begin{pmatrix}
	\theta 		&		\psi \\  
	-\psi 		&		\theta   
    \end{pmatrix},
\end{equation*}
which is, again, invertible unless $\theta$ and $\psi$ both vanish.  

In summary, imposing associativity of~$\wand$ restricts~$\bg$ to one of five possible standard forms, 
\begin{align*}
\bg &= (1,\, 0,\, 0,\, -1;\, 0,\, 1,\, 1,\, 0) \tag{C1}   		 \label{eqn:complex1}\\
\bg &= (1,\, 0,\, 0,\, 0;\, 0,\, 1,\, 1,\, 0)  \tag{C2}     	\label{eqn:complex2}\\
\bg &= (1,\, 0,\, 0,\, 0;\, 0,\, 0,\, 0,\, 1),  \tag{C3} 		 \label{eqn:complex3}\\
\intertext{and}
\bg &= (1,\, 0,\, 0,\, 0;\, 0,\, 1,\, 0,\, 0)  \tag{N1}  \label{eqn:nc1} \\ 
\bg &= (1,\, 0,\, 0,\, 0;\, 0,\, 0,\, 1,\, 0).  \tag{N2}  \label{eqn:nc2}  \\
\end{align*}
each of which, through Eq.~\eqref{eqn:wand-bilinear-form}, defines a way to multiply pairs.    The first three give complex multiplication~\eqref{eqn:complex1} followed by two variations thereof~(\ref{eqn:complex2} and~\ref{eqn:complex3}), and the last two give non-commutative multiplication~(\ref{eqn:nc1} and~\ref{eqn:nc2}). 

\subsection{Probability of a Sequence}
\label{sec:derivation2}

At this point, symmetry alone can take us no further in determining the precise form of the operator~$\wand$.  In order to make progress, we make use of the second part of our pair postulate, and  introduce a connection between the pair that represents a sequence and the probability associated with that same sequence.

We define the probability~$\prob{A}$ associated with sequence~$\sseq{A} = \llseq{\outcome_1}{\outcome_2}{\dots}{\outcome_n}$ as the probability of obtaining outcomes~$\outcome_2, \dots, \outcome_n$ conditional upon obtaining~$\outcome_1$,
\begin{equation} \label{eqn:def-of-probability-of-sequence}
\prob{A} = \Pr(\outcome_n, \outcome_{n-1}, \dots, \outcome_2 \,|\, \outcome_1).
\end{equation}
Following our pair postulate, we now require that~$\prob{A}$ is determined by the \weight,~$\pair{a}$, that represents sequence~$\sseq{A}$, so that, for any~$\pair{a}$,
\begin{equation}  \label{eqn:def-of-h}
\prob{A} = \hfunc(\pair{a}),
\end{equation}
where~$\hfunc$ is a continuous real-valued function that depends non-trivially on both real components of its argument~%
\footnote{This is necessary in order that both of the components of~$\pair{a}$ are relevant insofar as making experimental predictions is concerned.  If~$\hfunc$ were to depend upon only one component of~$\pair{a}$, the other component of the pair would be unused when computing probabilties.  Therefore, insofar as the making of physical predictions is concerned, the unused component could be deleted.  Thus, the pair representation could be reduced to a scalar representation without affecting predictive ability of the formalism.    Such a reduction would, however, be unacceptable as such a scalar representation would violate the requirement~(which underlies the pair postulate) that a measurement is only able to access one half of the degrees of freedom that are needed to describe a physical system.}. %
Our goal in this section is to determine the constraints imposed by probability theory on the form of~$\hfunc$ and, in the process of doing so, to show that only form~\eqref{eqn:complex1} can yield a form of~$\hfunc$ which meets our stated requirements.

\subsubsection{Probability Equation}   Consider the two sequences~$\sseq{A} = \seq{\outcome_1}{\outcome_2}$ and~$\sseq{B} = \seq{\outcome_2}{\outcome_3}$ of atomic outcomes.  Since outcome~$\outcome_2$ is the same in each,~$\sseq{C} = \sseq{A} \ser \sseq{B}$ is given by~$\sseq{C}  =  \lseq{\outcome_1}{\outcome_2}{\outcome_3}$.
The probability,~$\prob{C}$, associated with sequence~$\sseq{C}$ is given by
\begin{equation*}
\prob{C}   =  \Pr(\outcome_3, \outcome_2 \, | \, \outcome_1),
\end{equation*}
which, by the product rule of probability theory, can be rewritten as
\begin{equation*}
\prob{C}	=  \Pr(\outcome_3 \, | \, \outcome_2, \outcome_1)\Pr(\outcome_2  \,|\, \outcome_1).
\end{equation*}
Since~$\outcome_2$ is atomic, measurement~$\mment{M}{2}$~(with outcome~$\outcome_2$) establishes closure with respect to~$\mment{M}{3}$~(with outcome~$\outcome_3$), by definition overriding the earlier outcome~$\outcome_1$.
Therefore,  the probability of outcome~$\outcome_3$ is independent of~$\outcome_1$, and the above equation simplifies to
\begin{equation*}
\begin{aligned}
\prob{C}	&=   \Pr(\outcome_3  \, | \, \outcome_2)\Pr(\outcome_2 \, |\, \outcome_1) \\
		&=  \prob{B} \, \prob{A}.
\end{aligned}
\end{equation*}
Hence, for any~$\pair{a}, \pair{b}$, the function~$\hfunc$ must satisfy the equation
\begin{equation}  \label{eqn:h-product}
\hfunc(\pair{a} \wand \pair{b}) = \hfunc(\pair{a}) \,\hfunc(\pair{b}).
\end{equation}
Solving for the function~$\hfunc$ that satisfies this equation in each of the five forms of~$\bg$ given above,  we obtain
\begin{itemize}
\item[] \textrm{Case~\ref{eqn:complex1}}:    $\hfunc(\pair{a}) =  \left(a_1^2 + a_2^2\right)^{\alpha/2}$;  
\item[] \textrm{Case~\ref{eqn:complex2}}:  $\hfunc(\pair{a}) =  |a_1|^\alpha  e^{\beta a_2/a_1}$;
\item[] \textrm{Case~\ref{eqn:complex3}}:   $\hfunc(\pair{a}) =  |a_1|^\alpha |a_2|^\beta$;
\item[] \textrm{Case~\ref{eqn:nc1}}: $\hfunc(\pair{a}) = |a_1|^\alpha$;
\item[] \textrm{Case~\ref{eqn:nc2}}: $\hfunc(\pair{a}) = |a_1|^\alpha$; 
\end{itemize}
with~$\alpha, \beta$ real constants~(see Appendix~\ref{app:h-product-solutions}).
The solutions for~$\hfunc$ in the case of the two non-commutative forms~\eqref{eqn:nc1} and~\eqref{eqn:nc2} depend only on the first component of its argument.  That is not admissible, so those two forms are rejected.  
Of the five possible forms of~$\bg$, we are left with three:~\eqref{eqn:complex1}, \eqref{eqn:complex2} and~\eqref{eqn:complex3}.

\subsubsection{Reciprocity}

Suppose that the sequence~$\sseq{A} = \seq{m_1}{n_2}$ is obtained from an experiment where measurements~$\mmenta{M}$ and~$\mmenta{N}$ are performed at times~$t_1$ and~$t_2$, respectively.   Now consider the experiment where the measurements are performed in the reverse order, so that~$\mmenta{N}$ is performed at time~$t_1$, followed by~$\mmenta{M}$ at time~$t_2$, and suppose that the sequence obtained is $\revsseq{A} = \seq{n_2}{m_1}$, where the over-arrow symbolizes a unary operator acting on the sequence~$\sseq{A}$.

Suppose~$\mmenta{M}$ and~$\mmenta{N}$ are Stern-Gerlach measurements as in Sec.~\ref{sec:experimental-framework}.  Then, in the limit as~$t_2 \to t_1$, it follows from rotational symmetry and the empirical fact that the outcome probabilities of~$\mmenta{N}$ depend only on the magnitude of the angle between the axes of~$\mmenta{M}$ and~$\mmenta{N}$ that the probability~$\Pr(n_2|\, m_1)$ in the first experiment is equal to the probability~$\Pr(m_1 |\, n_2)$ in the second experiment.    Therefore, a relation is indicated between the pairs representing the sequences~$\sseq{A}$ and~$\revsseq{A}$.  
 
For our purpose, it is sufficient to assume that the \weight~$\revpair{a}$ that represents sequence~$\revsseq{A}$ is determined by the \weight~$\pair{a}$ that represents sequence~$\sseq{A}$ in the limit as~$t_2 \to t_1$,  so that
\begin{equation} 
\revpair{a} = \vectg{\pair{a}},
\end{equation}
where the reciprocity operator~$\vectgname{}$ is invertible~(since~$\vectg{\revpair{a}} = \pair{a}$), and is assumed continuous.    We shall assume that the above relation also holds more generally for sequences of arbitrary length.

Now, consider the sequences~$\sseq{A} = \lseq{\outcome_1}{\outcome_2}{\outcome_3}$ and~$\sseq{B} = \lseq{\outcome_1}{\outcome_2'}{\outcome_3}$, with~$\outcome_2 \neq \outcome_2'$, obtained from some experimental set-up, and the sequence,~$\sseq{C}$, that combines these in parallel, namely
\begin{equation}
\sseq{C} =  \sseq{A} \pll \sseq{B} 
			=   \lseq{\outcome_1}{\bubble{\outcome_2}{\outcome_2'}}{\outcome_{3}},
\end{equation}
and take the limit as the times,~$t_1, t_2$ and~$t_3$ of the respective measurements coincide.  The \weight\ that represents~$\revsseq{C}$ can be computed in two distinct ways, either as the \weight~$\vectg{\pair{c}}$, or as the \weight~$\vectg{\pair{a}} + \vectg{\pair{b}}$ that represents~$\revsseq{A} \pll \revsseq{B}$.  These two expressions must agree.  Therefore, for any~$\pair{a}$ and~$\pair{b}$,
\begin{equation} \label{eqn:rev-process-parallel}
\vectg{\pair{a} + \pair{b}} = \vectg{\pair{a}} + \vectg{\pair{b}},
\end{equation}
which implies linearity of~$\vectgname{}$,
\begin{equation} \label{eqn:R-linear}
\vectg{\pair{a}} = \pmat{R_1}{R_2}{R_3}{R_4} \cvpair{a}.
\end{equation}

Similarly, by considering two sequences~$\sseq{A}$ and~$\sseq{B}$ that can be combined in series to yield~$\sseq{C} = \sseq{A} \ser \sseq{B}$, and noting that~$\revsseq{C} = \revsseq{B} \ser \revsseq{A}$, one obtains 
\begin{equation} \label{eqn:rev-process-series}
 \vectg{\pair{a} \wand \pair{b}} = \vectg{\pair{b}} \wand \,\vectg{\pair{a}},
\end{equation}
which, for any selected form of multiplication~$\wand$, constrains the reciprocity coefficients~$R_1, \dots, R_4$.

\subsubsection{Repeated Measurements}

\begin{figure}[!h]
\includegraphics[width=5.5in]{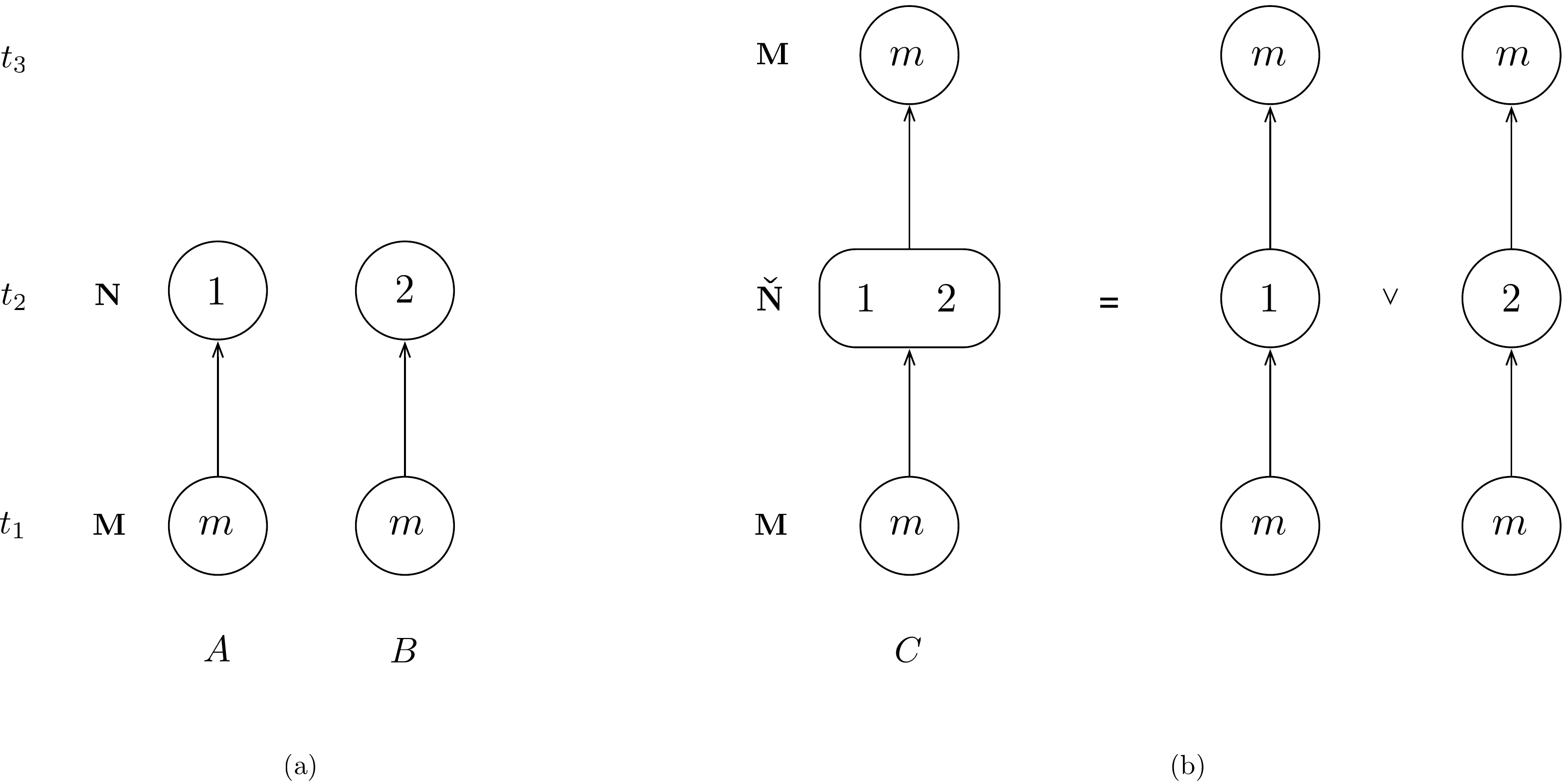}
\caption{\label{fig:repeated-measurements} Experiment~(a):~For given~$m$, the sequences~\sseq{A} and~\sseq{B} are mutually exclusive and exhaustive --- the outcome of~$\mmenta{N}$ must be~$1$ or~$2$, so that~$\prob{\sseq{A}} + \prob{\sseq{B}} = 1$.   Experiment~(b):~sequence~\sseq{C} has associated probability~$\prob{\sseq{C}} = 1$ in the limit as~$t_1, t_2$ and~$t_3$ coincide.}
\end{figure}

Consider an experiment in which measurements~$\mmenta{M}$ and~$\mmenta{N}$ are performed at times~$t_1$ and~$t_2$ respectively~(see Fig.~\ref{fig:repeated-measurements}a).   $\mmenta{N}$~allows only two atomic outcomes, $1$ or $2$.  Sequences~$\sseq{A} = \seq{\outcome}{1}$ and $\sseq{B} = \seq{\outcome}{2}$ have \weights~$\pair{a}$ and~$\pair{b}$, respectively.  Since either one outcome or the other occurs,~$\prob{\sseq{A}} + \prob{\sseq{B}} = 1$, so that
\begin{equation}  \label{eqn:sum-to-one}
\hfunc(\pair{a}) + \hfunc(\pair{b}) = 1.
\end{equation}
Now consider an experiment where measurement~$\mmenta{M}$ is performed at almost-coincident times~$t_1$ and~$t_3$, interleaved at intermediate time~$t_2$ by the trivial measurement~$\cmmenta{N}$ which has only one possible outcome~$\bubble{1}{2}$~(see Fig.~\ref{fig:repeated-measurements}b).     
The sequence~$\lseq{\outcome}{\bubble{1}{2}}{\outcome}$ can be written as
\begin{equation*}
    \sseq{C} =  \lseq{\outcome}{1}{\outcome}  \pll \lseq{\outcome}{2}{\outcome}
\end{equation*}
and, because the time-offsets are negligible, we also have that
\begin{equation*}
  \lseq{\outcome}{1}{\outcome} = \seq{\outcome}{1} \ser \, \seq{1}{\outcome} = \sseq{A} \ser \revsseq{A}
    \quad\quad\text{and}\quad\quad
  \lseq{\outcome}{2}{\outcome} = \seq{\outcome}{2} \ser \, \seq{2}{\outcome} = \sseq{B} \ser \revsseq{B}.
\end{equation*}
Therefore the pair that represents~$\sseq{C}$ is
\begin{equation}
\pair{c} = \left( \pair{a} \wand \vectg{\pair{a}} \right) + \left( \pair{b} \wand \vectg{\pair{b}} \right).
\end{equation}

Now, the intermediate measurement,~$\cmmenta{N}$, is trivial in that it only registers that a physical system is detected in the measuring device at time~$t_2$, but demonstrably does not affect the outcome probabilities of subsequent measurements performed upon the system.   As measurement~$\mmenta{M}$ is repeatable~(see Sec.~\ref{sec:experimental-framework}), it follows that, in the limit as~$t _1, t_2$ and~$t_3$ coincide,
\begin{equation} \label{eqn:prob-limit}
\cprob{a} + \cprob{b}  = 1
    \quad\quad\Longrightarrow\quad\quad
\cprob{c}  = 1.
\end{equation} 
We are now in a position to eliminate forms~\eqref{eqn:complex2} and~\eqref{eqn:complex3}, leaving~\eqref{eqn:complex1} together with the specific form of~$\hfunc$.

\subparagraph{Form~\eqref{eqn:complex2}.}   Multiplication is via~$\bg =(1,\, 0,\, 0,\, 0;\, 0,\, 1,\, 1,\, 0)$, with~$\hfunc(\pair{x}) = |x_1|^\alpha e^{\beta x_2/x_1}$. 

On substituting the linear form of Eq.~\eqref{eqn:R-linear} into Eq.~\eqref{eqn:rev-process-series}, one finds that the only non-trivial reciprocity operator is
\begin{equation*}
\vectg{\pair{a}} = \pv{a_1}{0}.
\end{equation*}
This is not invertible, which at once eliminates~\eqref{eqn:complex2}.

\subparagraph{Form~\eqref{eqn:complex3}.}   Multiplication is via~$\bg =(1,\, 0,\, 0,\, 0;\, 0,\, 0,\, 0,\, 1)$, with~$\hfunc(\pair{x}) = |x_1|^\alpha |x_2|^\beta$.

On substituting the linear form of Eq.~\eqref{eqn:R-linear} into Eq.~\eqref{eqn:rev-process-series}, one finds that the only non-trivial reciprocity operators are
\begin{equation*} 
\vectg{\pair{a}} = \pv{a_2}{a_1}
\quad\quad\text{and}\quad\quad
\vectg{\pair{a}} = \cvpair{a}.
\end{equation*}
These are invertible, as required.   

Choosing~$\vectg{\pair{a}} = \lpair{a _2}{a _1}$ makes~$\pair{c} =\lpair{a_1a_2 + b_1b_2}{a_1a_2 + b_1b_2}$ so that
 Eq.~\eqref{eqn:prob-limit} reads
\begin{equation*}
|a _1|^\alpha |a_2|^\beta + |b_1|^\alpha |b_2|^\beta = 1  \quad\quad\Longrightarrow\quad \quad |a_1a_2 + b_1b_2|^{\alpha + \beta} =1.
\end{equation*}
The special case~$b_1b_2 = -a_1 a_2$ can satisfy the left condition while contradicting the right, thereby disproving this choice.

The other choice is~$\vectg{\pair{a}} = \cpair{a}$, for which~$\pair{c} = \lpair{a_1^2 + b_1^2}{a _2^2 + b _2^2}$, so that 
 Eq.~\eqref{eqn:prob-limit} reads
\begin{equation*}
|a _1|^\alpha |a_2|^\beta + |b_1|^\alpha |b_2|^\beta = 1  \quad\quad\Longrightarrow\quad \quad \left(a_1^2 + b_1^2 \right)^\alpha \left(a_2^2 + b_2^2 \right)^\beta =1.
\end{equation*}
The special case~$a_1 = b_2 = rt$ and~$a_2 = b_1 = r/t$ with~$r, t \neq 0$ reduces this to the identity
\begin{equation*}
 \left( t^{\alpha - \beta}  + t^{\beta - \alpha} \right)^2  = \left( t^2 + t^{-2} \right)^{\alpha + \beta},
\end{equation*}
valid for arbitrary~$t$.  This requires either~$\alpha = 2$ with~$\beta =0$, or~$\alpha =0$ with~$\beta = 2$.  But~$\alpha =0$ makes~$\hfunc(\pair{a})$ independent of~$a_1$, and~$\beta = 0$ makes~$\hfunc(\pair{a})$ independent of~$a_2$, whereas we require~$\hfunc(\pair{a})$ to depend on both arguments.  Hence, this choice too is disproved, which eliminates~\eqref{eqn:complex3}.

\subparagraph{Form~\eqref{eqn:complex1}.}   Multiplication is via~$\bg = (1,\, 0,\, 0,\, -1;\, 0,\, 1,\, 1,\, 0)$, with~$\hfunc(\pair{x}) = \left(x_1^2 + x_2^2 \right)^{\alpha/2}$.   
On substituting the linear form of Eq.~\eqref{eqn:R-linear} into Eq.~\eqref{eqn:rev-process-series}, one finds that the only non-trivial reciprocity operators are
\begin{equation*}
\vectg{\pair{a}} = \cvpair{a}  
\quad\quad\text{and}\quad\quad
\vectg{\pair{a}} = \pv{a_1}{- a_2}.
\end{equation*}
These are invertible, as required.   

Choosing~$\vectg{\pair{a}} = \cpair{a}$ makes~$\pair{c} = \lpair{a_1^2 - a_2^2 + b_1^2 - b_2^2}{2a_1a_2 + 2b_1b_2}$, so that
 Eq.~\eqref{eqn:prob-limit} reads
\begin{equation*}
 \left(a_1^2 + a_2^2\right)^{\alpha/2} + \left(b_1^2 + b_2^2\right)^{\alpha/2} =1  \quad\quad\Longrightarrow\quad \quad  \left[ \left(a_1^2 - a_2^2 + b_1^2 - b_2^2\right)^2 + 4\left(a_1a_2 + b_1b_2\right)^2\right]^{\alpha/2}  =1.
\end{equation*}
The special case~$b_1 = a_2, b_2 = -a_1$ can satisfy the left condition while contradicting the right, thereby disproving this choice.

The other choice is~$\vectg{\pair{a}} = \lpair{a_1}{-a_2}$, for which~$\pair{c} = \lpair{a_1^2 + a_2^2 +  b_1^2 + b_2^2}{0}$, so that
 Eq.~\eqref{eqn:prob-limit} reads
\begin{equation*}
 \left(a_1^2 + a_2^2\right)^{\alpha/2} + \left(b_1^2 + b_2^2\right)^{\alpha/2} =1  \quad\quad\Longrightarrow\quad \quad  \left(a_1^2 + a_2^2 + b_1^2 + b_2^2\right)^\alpha  =1.
\end{equation*}
This requires~$\alpha =2$, and this setting gives an admissible solution.  Hence 
\begin{equation}  \label{eqn:Born}
\hfunc(\pair{x}) = x_1^2 + x_2^2.
\end{equation}
We are left with just this one solution.

\subsection{Summary}

In order to combine sequences in parallel, we have the sum rule of Eq.~\eqref{eqn:weight-sum-rule}, 
\begin{equation*}
\cvpair{a} \wor \cvpair{b} = \pv{a_1 + b_1}{a_2 + b_2},
\end{equation*}
which we recognize as complex addition.   
In order to combine sequences in series, from Eq.~\eqref{eqn:wand-bilinear-form} with~$\bg$ given by the surviving form~\eqref{eqn:complex1}, we have
\begin{equation*}
\cvpair{a} \wand \cvpair{b} = \pv{a_1b_1 - a_2b_2}{a_1b_2 + a_2b_1},
\end{equation*}
which we recognize as complex multiplication.   
Hence the number pairs~$\pair{a}, \pair{b}, \dots$ behave as complex numbers, combining according to the rules of complex arithmetic.  For the probability associated with a sequence, form~\eqref{eqn:complex1} gives Eq.~\eqref{eqn:Born}, namely
\begin{equation*}
\hfunc(\pair{x}) = x_1^2 + x_2^2.
\end{equation*}
These are Feynman's rules.

\section{Discussion} \label{sec:discussion}

In this paper, we have shown that the concept of complementarity, regarded by Bohr as one of the most fundamental lessons of quantum phenomena for our physical world-view~\cite{Bohr28,Bohr-complementarity,Bohr-Sci-Human-Knowledge-ch1}, can be used to derive quantum theory.  
In particular, once complementarity is postulated in the minimalist form that there are \emph{two} real degrees of freedom associated with each sequence~(or `path'), but that one can only access \emph{one} real-valued function of these degrees of freedom in a given experiment,  the complex arithmetic of the quantum formalism emerges naturally.

It is also interesting to consider which non-classical features the derivation does \emph{not} use.   It has been suggested~\cite{Popescu-Rohrlich97} that the quantum formalism may owe at least a significant part of its structure to the fact that quantum theory permits non-locality and no-signaling to peacefully coexist, and a number of recent reconstructive approaches~\cite{Clifton-Bub-Halvorson03, Hardy01a, Hardy01b} rely upon postulates that concern the behavior of physically separated sub-systems.  However, the derivation we present here takes place without making reference to more than one physical system, and thus demonstrates that features such as non-locality and no-signalling are \emph{not}, in fact, essential to an understanding of the structure of the quantum formalism.   That is, from the point of view of the present derivation, features such as non-locality and no-signalling are not fundamental, but secondary.

The derivation also illuminates the nature of the relationship between the quantum formalism and the fundamental concepts of classical physics.  The original formulations of quantum theory both make explicit use of the structure of classical physics\,---\,Schroedinger's derivation~\cite{Schroedinger26} was based directly on de Broglie's wave-particle duality~(itself based on the classical models of waves and particles), while Heisenberg's derivation took the classical model of electromagnetic radiation from atoms as its point of departure~\cite{Heisenberg25}.  Hence, both of these formulations presuppose the gamut of fundamental classical concepts such as space, time, matter, energy, and momentum.   Therefore, the question naturally arises as to whether this is an historical accident and the quantum formalism is, in fact, prior to these concepts.

In contrast to Schroedinger's and Heisenberg's formulations, the derivation presented here makes no explicit reference to the classical concepts of space, energy, and momentum.  The only aspect of the concept of time which has been assumed is the time-ordering of events; in particular, quantification of time~(on the real number line) plays no role.    Hence, the derivation shows that the core of the quantum formalism is a self-contained theoretical structure that makes minimal use of the fundamental concepts of classical physics.   Of particular relevance to the programme of quantum gravity, the derivation clearly suggests that the Feynman rules are logically prior to the structure of space or, more generally, of spacetime.   

Since the formulation of quantum theory, numerous proposals have been made on how the formalism could be modified in various ways, such as allowing non-linear continuous transformations~\cite{Weinberg89a,Weinberg89b} and modifying the formalism to use quaternions~\cite{Finkelstein-quaternionic-QM} or $p$-adic numbers~\cite{Meurice-p-adic}.   Although these proposed modifications may be mathematical plausible, they are rather \emph{ad hoc} from a physical point of view since they are not clearly driven by physical facts or principles, and furthermore are difficult to subject to experimental tests.    The derivation of Feynman's rules given here provides a natural framework within which such proposals can be systematically studied. 

Feynman's rules do not exhaust the content of the standard quantum formalism.  For example, once translated into the von Neumann-Dirac state picture~(see, for example~\cite{Feynman48,Caticha98b}), the Feynman rules imply that state evolution is linear, but do not imply that it is unitary.  Therefore, one must appeal to additional arguments~(such as Wigner's theorem) to establish unitarity.   It has been shown elsewhere that, given unitarity, the remaining standard structure\,---\,the tensor product rule, the representation of reproducible measurements by means of Hermitian operators, the general form of the temporal evolution operator, and the explicit forms of commonly used measurement operators\,---\,can all be systematically reconstructed~\cite{Goyal-QT2c,Caticha98b}.  Hence, the present derivation  provides a sound basis for the reconstruction of the entirety of the standard von Neumann-Dirac quantum formalism for finite-dimensional quantum systems.

\section{Acknowledgements}

Philip Goyal would like to thank Janos Acz\'{e}l, Bob Coecke, Yiton Fu, and Luca Mana for very helpful discussions, and Perimeter Institute for excellent institutional support.   Research at Perimeter Institute is supported in part by the Government of Canada through NSERC and by the Province of Ontario through MEDT.     Kevin Knuth would like to thank Ariel Caticha and Carlos Rodr\'iguez for many insightful discussions.  John Skilling thanks the Perimeter Institute for hospitality.

%%%%%%%%%%%%%%%%% Appendix   %%%%%%%%%%%%%%%%%
\appendix

\section{Solution of $\wand$-Associativity Equation} \label{app:wand-associativity}

Substituting Eq.~\eqref{eqn:wand-bilinear-form} into Eq.~\eqref{eqn:wand-associativity}, and equating the first and second components, respectively, yields the following two equations:
\begin{multline*}
(\g_1^2 + \g_3 \g_5) a_1 b_1 c_1 + (\g_1 \g_2 + \g_4 \g_5) a_1 b_1 c_2
      + (\g_1 \g_2 + \g_3 \g_6) a_1 b_2 c_1 + (\g_2^2        + \g_4 \g_6) a_1 b_2 c_2    \\
       + (\g_1 \g_3 + \g_3 \g_7) a_2 b_1 c_1 + (\g_2 \g_3 + \g_4 \g_7) a_2 b_1 c_2
      + (\g_1 \g_4 + \g_3 \g_8) a_2 b_2 c_1 + (\g_2 \g_4 + \g_4 \g_8) a_2 b_2 c_2       \\
 =     (\g_1^2        + \g_2 \g_5) a_1 b_1 c_1 + (\g_1 \g_2 + \g_2 \g_6) a_1 b_1 c_2
      + (\g_1 \g_3 + \g_2 \g_7) a_1 b_2 c_1 + (\g_1 \g_4 + \g_2 \g_8) a_1 b_2 c_2   \\
	+ (\g_1 \g_3 + \g_4 \g_5) a_2 b_1 c_1 + (\g_2 \g_3 + \g_4 \g_6) a_2 b_1 c_2
      + (\g_3^2        + \g_4 \g_7) a_2 b_2 c_1 + (\g_3 \g_4 + \g_4 \g_8) a_2 b_2 c_2
\end{multline*}
and
\begin{multline*}
        (\g_1 \g_5 + \g_5 \g_7) a_1 b_1 c_1 + (\g_1 \g_6 + \g_5 \g_8) a_1 b_1 c_2
      + (\g_2 \g_5 + \g_6 \g_7) a_1 b_2 c_1 + (\g_2 \g_6 + \g_6 \g_8) a_1 b_2 c_2 \\
	+ (\g_3 \g_5 + \g_7^2       ) a_2 b_1 c_1 + (\g_3 \g_6 + \g_7 \g_8) a_2 b_1 c_2
      + (\g_4 \g_5 + \g_7 \g_8) a_2 b_2 c_1 + (\g_4 \g_6 + \g_8^2       ) a_2 b_2 c_2 \\
      =     (\g_1 \g_5 + \g_5 \g_6) a_1 b_1 c_1 + (\g_2 \g_5 + \g_6^2       ) a_1 b_1 c_2
      + (\g_3 \g_5 + \g_6 \g_7) a_1 b_2 c_1 + (\g_4 \g_5 + \g_6 \g_8) a_1 b_2 c_2 \\
	+ (\g_1 \g_7 + \g_5 \g_8) a_2 b_1 c_1 + (\g_2 \g_7 + \g_6 \g_8) a_2 b_1 c_2
      + (\g_3 \g_7 + \g_7 \g_8) a_2 b_2 c_1 + (\g_4 \g_7 + \g_8^2       ) a_2 b_2 c_2.
\end{multline*}
These equations must hold for any $a_1, a_2, b_1, b_2, c_1, c_2$.    Equating coefficients, we get sixteen equations which, upon factorization and removal of redundant equations, reduce to twelve equations,
\begin{align}
	   \g_2 \g_6 			 &= \g_4 \g_5                              \label{eqn:g1}\\
          \g_3 \g_7 					   &= \g_4 \g_5                             \label{eqn:g2} \\
          \g_4 (\g_2 - \g_3)  &= 0                           						\label{eqn:g3}\\
          \g_4 (\g_6 - \g_7) &= 0                           						\label{eqn:g4}\\
          \g_5 (\g_2 - \g_3) &= 0                           						\label{eqn:g5}		\\
          \g_5 (\g_6 - \g_7) &= 0                           						\label{eqn:g6}\\
          \g_2 (\g_1 - \g_7) &= \g_3 (\g_1 - \g_6)    						\label{eqn:g7}\\
          \g_4 (\g_1 - \g_7) &= \g_3 (\g_3 - \g_8)    						\label{eqn:g8}\\
          \g_7 (\g_1 - \g_7) &= \g_5 (\g_3 - \g_8)    						\label{eqn:g9}\\
          \g_7 (\g_2 - \g_8) &= \g_6 (\g_3 - \g_8)    						\label{eqn:g10}\\
          \g_5 (\g_2 - \g_8) &= \g_6 (\g_1 - \g_6)    						\label{eqn:g11}\\
          \g_2 (\g_2 - \g_8) &= \g_4 (\g_1 - \g_6).							\label{eqn:g12}
\end{align}

To solve the above equations, we select the nature of $\g_6$ and $\g_7$, choosing from the cases~$\g_6 = \g_7 \ne 0$, or $\g_6 \ne \g_7$, or $\g_6 = \g_7 = 0$.

\subsection{Case~$\g_6 = \g_7 \ne 0$.}
In this case, the first two equations give
\begin{equation*}
    \g_2 = \g_3 = \frac{\g_4 \g_5}{\g_6},
\end{equation*}
while the remainder reduce to
\begin{equation*}
\begin{aligned}
 \g_4 (\g_1 - \g_6) &= \g_2 (\g_2 - \g_8) \\
 \g_6 (\g_1 - \g_6) &= \g_5 (\g_2 - \g_8),
\end{aligned}
\end{equation*}
which both read
\begin{equation*}
    \g_1 = \g_6 + \frac{\g_5 (\g_2 - \g_8)}{\g_6}.
\end{equation*}
Therefore, 
\begin{equation*}
    \bg = \left( \g_6 + \frac{\g_5}{\g_6} \Big( \frac{\g_4\g_5}{\g_6} - \g_8\Big),
                    { \frac{\g_4\g_5}{\g_6}}, { \frac{\g_4\g_5}{\g_6}}, \g_4; \g_5, \g_6, \g_6, \g_8 \right),  
\end{equation*}
which we can write in the more symmetric form
\begin{equation} \label{eqn:soln-comm}
\bg = (\theta - \psi\epsilon,\, \phi\epsilon,\, \phi\epsilon,\, \phi;\  \theta\epsilon,\, \theta,\, \theta,\, \psi + \phi\epsilon),    \tag{A}
\end{equation}
with real constants~$\theta, \phi, \psi, \epsilon$.

\subsection{Case~$\g_6 \neq \g_7$.}

In this case, Eqs.~\eqref{eqn:g4} and~\eqref{eqn:g6} give~$\g_4 = \g_5 = 0$, and the remaining equations are
\begin{align}
\gamma_2 \gamma_6 						&=   0                         							 \tag{\ref{eqn:g1}$'$}     		\label{eqn:g1d}   \\
\gamma_3 \gamma_7 						&=   0                          						 \tag{\ref{eqn:g2}$'$}     \label{eqn:g2d}   \\
\gamma_2 (\gamma_1 - \gamma_7) &=  \gamma_3 (\gamma_1 - \gamma_6)   \tag{\ref{eqn:g7}$'$}     \label{eqn:g7d}   \\         
\gamma_3 (\gamma_3 - \gamma_8) &= 0                                \tag{\ref{eqn:g8}$'$}     \label{eqn:g8d}   \\
          \gamma_7 (\gamma_1 - \gamma_7) &= 0                     \tag{\ref{eqn:g9}$'$}     \label{eqn:g9d}   \\          \gamma_7 (\gamma_2 - \gamma_8) &= \gamma_6 (\gamma_3 - \gamma_8)   \tag{\ref{eqn:g10}$'$}     \label{eqn:g10d}   \\
\gamma_6 (\gamma_1 - \gamma_6) &= 0                                \tag{\ref{eqn:g11}$'$}     \label{eqn:g11d}   \\          \gamma_2 (\gamma_2 - \gamma_8) &= 0.                               \tag{\ref{eqn:g12}$'$}     \label{eqn:g12d}   
\end{align}
If both $\gamma_6$ and $\gamma_7$ are non-zero, then Eqs.~\eqref{eqn:g9d} and~\eqref{eqn:g11d} imply $\gamma_6 = \gamma_7\ (= \gamma_1)$, contrary to assumption.  Hence exactly one of them must be zero.
Suppose, then, that~$\gamma_6 = 0$, with $\gamma_7 \ne 0$.    Then $\gamma_3 = 0$, $\gamma_1 = \gamma_7$, and $\gamma_2 = \gamma_8$, giving
\begin{equation}
    \bg = (\gamma_1,\, \gamma_2,\, 0,\, 0;\, 0,\, 0,\, \gamma_1,\, \gamma_2). \tag{B} \label{eqn:soln-nc1}
\end{equation}
Similarly, suppose that~$\gamma_7 = 0$, with $\gamma_6 \ne 0$.  Then~$\gamma_2 = 0$, $\gamma_1 = \gamma_6$, and~$\gamma_3 = \gamma_8$, giving
\begin{equation}
    \bg = (\gamma_1,\, 0,\, \gamma_3,\, 0;\, 0,\, \gamma_1,\, 0,\, \gamma_3).    \tag{C} \label{eqn:soln-nc2}
\end{equation}

\subsection{Case~$\g_6 = \g_7 =0$.}

Before considering this case, we consider the solution of Eqs.~\eqref{eqn:g1}--\eqref{eqn:g12} with respect to the nature of~$\g_2$ and~$\g_3$, choosing from $\gamma_2 = \gamma_3 \ne 0$, or $\gamma_2 \ne \gamma_3$, or $\gamma_2 = \gamma_3 = 0$.   The treatment mirrors that of $\gamma_6$ and $\gamma_7$ just given.   The first choice repeats solution~\eqref{eqn:soln-comm} and the second repeats solutions~\eqref{eqn:soln-nc1} and~\eqref{eqn:soln-nc2}.  

All that remains is $\gamma_2 = \gamma_3 = 0$, which we only need analyze in the context of~$\gamma_6 = \gamma_7 = 0$.    The surviving equations reduce to  
\begin{equation*}
    \gamma_1 \gamma_4 = \gamma_4 \gamma_5 = \gamma_5 \gamma_8 = 0,
\end{equation*}
whose solutions
\begin{equation*}
\begin{aligned}
\bg &= (\gamma_1,\, 0,\, 0,\, 0;\, \gamma_5,\, 0,\, 0,\, 0) \\
\bg &= (\gamma_1,\, 0,\, 0,\, 0;\, 0,\, 0,\, 0,\, \gamma_8) \\
\bg &= (0,\, 0,\, 0,\, \gamma_4;\, 0,\, 0,\, 0,\, \gamma_8) 
\end{aligned}
\end{equation*}
are special or limiting cases of solution~\eqref{eqn:soln-comm}.  

Hence, the possible solutions for~$\bg$ are the commutative solution~\eqref{eqn:soln-comm} and the two non-commutative solutions~\eqref{eqn:soln-nc1} and~\eqref{eqn:soln-nc2}.

\section{Solutions of Probability Equation}
\label{app:h-product-solutions}

We solve the probability equation, Eq.~\eqref{eqn:h-product}, for each of the five standard forms of~$\bg$ with the aid of two of Cauchy's standard functional equations
\begin{align*}
  f(xy) = f(x)\,f(y)     \quad\quad\text{and}\quad\quad       f(x+y) = f(x)\,f(y).
\end{align*}
We quote as needed~\cite{Aczel-lectures-functional-equations-ch2} their continuous solutions, respectively
\begin{align*}
  f(x) = |x|^\alpha  \quad\quad\text{and}\quad\quad    f(x) = e^{\beta x}. 
\end{align*}

\subsection*{Form~\eqref{eqn:complex1}:~$\bg = (1,\,0,\,0,\,-1; \,0,\,1,\,1,\,0)$.}
Explicitly,~Eq.~\eqref{eqn:h-product} reads
\begin{equation}  \label{eqn:h-product1}
\hfunc(a_1b_1 - a_2b_2, a_1b_2 + a_2b_1) = \hfunc(a_1, a_2) \,\hfunc(b_1, b_2) 
\end{equation}
for arbitrary $a_1$, $a_2$, $b_1$, $b_2$.   Change variables by setting $a_1 = r\cos\theta$, $a_2 = r\sin\theta$, $b_1 = s\cos\phi$, $b_2 = s\sin\phi$, with~$r, s \geq 0$, to obtain
\begin{equation}  \label{eqn:h-product1-angle-form}
    \hfunc\left(rs\cos(\theta+\phi), rs\sin(\theta+\phi)\right) = \hfunc(r\cos\theta, r\sin\theta) \,\hfunc(s\cos\phi, s\sin\phi).
\end{equation}
In case~$r=s=1$, this takes the form
\[
    f(\theta + \phi) = f(\theta) f(\phi)
\]
with~$f(\psi) \equiv \hfunc\left( \cos\psi, \sin\psi\right)$, which has the solution~$f(\psi) = e^{\beta\psi}$.  Since~$f(\psi + 2\pi) = f(\psi)$,~$\beta=0$, so that
\begin{equation*} \label{eqn:p-on-circle}
f(\psi) = p(\cos\psi, \sin\psi) = 1.
\end{equation*}
Using this in Eq.~\eqref{eqn:h-product1-angle-form} with~$s=1$ and~$\theta = 0$, we obtain
\begin{equation} \label{eqn:h-product1-moving-circle}
 \hfunc\left(r\cos\phi, r\sin\phi \right)  = \hfunc(r,0),
\end{equation}
which reduces Eq.~\eqref{eqn:h-product1-angle-form} to
\[
p(rs, 0) = p(r, 0) \, p(s, 0).
\]
This has solution~$p(t,0) = t^\alpha$.  Hence, from Eq.~\eqref{eqn:h-product1-moving-circle},~$p(r\cos\phi, r\sin\phi) = r^\alpha$.  Re-writing the arguments of~$\hfunc$ yields
\begin{equation}  \label{eqn:product1-solution}
\hfunc(x_1, x_2) = \left(x_1^2 + x_2^2 \right)^{\alpha/2}.
\end{equation}
This satisfies Eq.~\eqref{eqn:h-product1}, so is the general solution.

\subsection*{Form~\eqref{eqn:complex2}:~$\bg = (1,\,0,\,0,\,0;\,0,\,1,\,1,\,0)$.}

Explicitly, Eq.~\eqref{eqn:h-product} reads
\begin{equation} \label{eqn:h-product2}
    p(a_1 b_1, a_1 b_2 + a_2 b_1) = p(a_1, a_2) \, p(b_1, b_2)   
\end{equation}
for arbitrary $a_1$, $a_2$, $b_1$, $b_2$. 
In case $a_1 = b_1 = 1$, this reduces to $p(1, a_2 + b_2) = p(1,a_2)\,p(1,b_2)$, whose solution is
\begin{equation}
    p(1,x_2) = e^{\beta x_2}. \label{eqn:h-product2-B6}
\end{equation}
In case $a_2 = b_2 = 0$, Eq.~\eqref{eqn:h-product2} reduces to $p(a_1 b_1, 0) = p(a_1, 0)\,p(b_1, 0)$, whose solution is
\begin{equation}
    p(x_1, 0) = |x_1|^\alpha. \label{eqn:h-product2-B7}
\end{equation}
In case $a_1 = b_2 = 1$, $a_2 = -1/b_1$ with $b_1\ne 0$, Eq.~\eqref{eqn:h-product2} reduces to $p(b_1, 0) = p(1, -1/b_1)\,p(b_1, 1)$.
Using Eq.~\eqref{eqn:h-product2-B6} and Eq.~\eqref{eqn:h-product2-B7}, this gives
\begin{equation}
    p(b_1, 1) = |b_1|^\alpha e^{\beta/b_1}. \label{eqn:h-product2-B8}
\end{equation}
In case $a_1 = b_2 = 1$, Eq.~\eqref{eqn:h-product2} reduces to $p(b_1, 1 + a_2 b_1) = p(1, a_2)\,p(b_1, 1)$.
Using Eq.~\eqref{eqn:h-product2-B6} and Eq.~\eqref{eqn:h-product2-B8}, this gives
\begin{equation*}
    p(b_1, 1 + a_2 b_1) = |b_1|^\alpha e^{\beta(1 + a_2 b_1)/b_1}
\end{equation*}
from which the solution can be read off as
\begin{equation}
p(x_1, x_2) = |x_1|^\alpha e^{\beta x_2/x_1}.
\end{equation}
This satisfies Eq.~\eqref{eqn:h-product2}, so is the general solution.

%%%%%%%%%%%%% case A3 %%%%%%%%%%%%%%%%%%%%%%%

\subsection*{Form~\eqref{eqn:complex3}:~$\bg = (1,\,0,\,0,\,0;\, 0,\,0,\,0,\,1)$.}
Explicitly, Eq.~\eqref{eqn:h-product} reads
\begin{equation} \label{eqn:h-product3} 
p(a_1 b_1, a_2 b_2) = p(a_1, a_2) \, p(b_1, b_2)
\end{equation}
for arbitrary $a_1$, $a_2$, $b_1$, $b_2$.
In case~$a_2=b_2 =1$, this reduces to
\[
p(a_1b_1, 1) = p(a_1, 1) \, p(b_1, 1),
\]
whose solution is
\[
p(x_1, 1) = |x_1|^\alpha.
\]
Similarly, by considering case~$a_1 = b_1 =1$, we obtain~$p(1, x_2) = |x_2|^\beta$.  Using these special solutions in Eq.~\eqref{eqn:h-product3} with~$(a_1, a_2) = (x_1, 1)$ and~$(b_1,  b_2) = (1, x_2)$ yields
\begin{equation}
p(x_1, x_2) = |x_1|^\alpha |x_2|^\beta.
\end{equation}
This satisfies Eq.~\eqref{eqn:h-product3}, so is the general solution.

%%%%%%%%%%%%% case B1 %%%%%%%%%%%%%%%%%%%%%%%

\subsection*{Form~\eqref{eqn:nc1}:~$\bg = (1,\,0,\,0,\,0; \,0,\,1,\,0,\,0)$.}

Explicitly, Eq.~\eqref{eqn:h-product} reads
\begin{equation} \label{eqn:h-product4}
p(a_1 b_1, a_1 b_2) = p(a_1, a_2) \, p(b_1, b_2)
\end{equation}
for arbitrary $a_1$, $a_2$, $b_1$, $b_2$.    The left side is independent of $a_2$, so $p$ cannot depend on its second argument ($a_2$ on the right).  Hence,~$p(x_1, x_2) = f(x_1)$. Eq.~\eqref{eqn:h-product4} thus reduces to~$f(a_1b_1) = f(a_1)f(b_1)$, whose solution is~$f(x_1) = |x_1|^\alpha$.  Hence, the solution of Eq.~\eqref{eqn:h-product4} is
\begin{equation}
    p(x_1,x_2) = |x_1|^\alpha.
\end{equation}
This satisfies Eq.~\eqref{eqn:h-product4}, so is the general solution.

\subsection*{Form~\eqref{eqn:nc2}:~$\bg = (1,\,0,\,0,\,0; \,0,\,0,\,1,\,0)$.}

Explicitly, Eq.~\eqref{eqn:h-product} reads
\begin{equation} \label{eqn:h-product5}
    p(a_1b_1, a_2b_1) = p(a_1, a_2) \, p(b_1, b_2)  
\end{equation}
for arbitrary $a_1$, $a_2$, $b_1$, $b_2$.
Arguing as above, the left side is independent of $b_2$, so $p$ cannot depend on its second argument ($b_2$ on the right).
Hence the solution of the above equation is also
\begin{equation*} 
    p(x_1,x_2) = |x_1|^\alpha.
\end{equation*}
This satisfies Eq.~\eqref{eqn:h-product5}, so is the general solution.

%\bibliography{references}

\end{document}